\documentclass[twocolumn]{svjour3}
\usepackage[utf8]{inputenc}

\usepackage{graphicx}
\usepackage{xcolor} 
\usepackage{amsmath}
 \journalname{Granular Matter}
 
\begin{document}

\title{The intertwined roles of particle shape and surface roughness in controlling the shear strength of a granular material}

\author{Kieran A. Murphy         \and
        Arthur K. MacKeith       \and
        Leah K. Roth            \and
        Heinrich M. Jaeger  
}
\institute{K.A. Murphy \and
		A.K. MacKeith \and
		L.K. Roth \and
		H. M. Jaeger\\
              Department of Physics and James Franck Institute \\
              The University of Chicago \\
              Chicago, Illinois 60637 USA \\
              \email{jaeger@uchicago.edu}    
}

\date{Received: date / Accepted: date}

\maketitle

\begin{abstract}
Both the shape of individual particles and their surface properties contribute to the strength of a granular material under shear. 
Here we show the degree to which these two aspects can be intertwined. 
In experiments on assemblies of 3D printed, convex lens-shaped particles, we measure the stress-strain response under repeated compressive loading and find that the aggregate's shear strength falls rapidly when compared to other particle shapes.
We probe the granular material at mm-scales with X-ray computed tomography and $\mu$m-scales with high-resolution surface metrology to look for the cause of the degradation.
We find that wear due to accumulated deformation smooths out the lens surfaces in a controlled and systematic manner that correlates with a significant loss of shear strength observed for the assembly as a whole. 
The sensitivity of lenses to changes in surface properties contrasts with results for assemblies of 3D printed tetrahedra and spheres, which under the same load cycling are found to exhibit only minor degradation in strength.  
This case study provides insight into the relationship between particle shape, surface wear, and the overall material response, and suggests new strategies when designing a granular material with desired evolution of properties under repeated deformation.

\end{abstract}

\section{Introduction}

In a granular material, particle shape changes the manner in which surfaces meet to support stress, whether by edges, corners, or surfaces with different radii of curvature. 
Shape also determines the resistance particles experience toward reconfiguring along rotational and translational degrees of freedom.
As a result, particle shape greatly affects the bulk properties of granular materials \cite{ThanasiShapes,StannariusShape2013,GravishStaples2012,Zpaper}.
Aggregates of angular particles have higher shear strength than those of rounded particles \cite{AnthonyParticleCharacteristics2005,Santamarinasoilshape2004,AzemaPentagons2007}, and platy particles tend to plastically deform with larger stress fluctuations than compact shapes \cite{StressFlucsKM}.

\begin{figure*}
\centering
  \includegraphics[width=2\columnwidth]{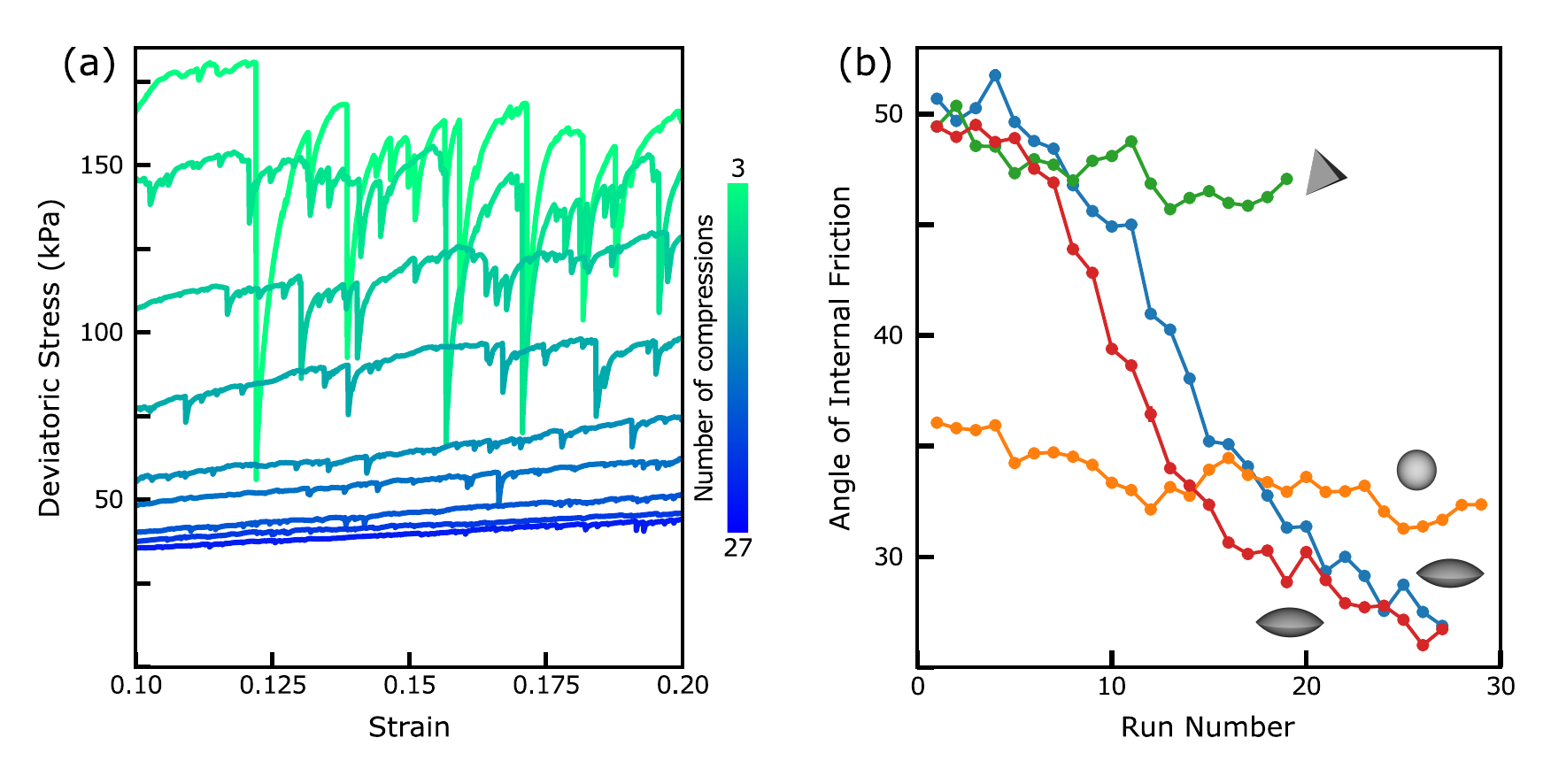}
\caption{\textbf{(a)} Raw stress-strain data for every third uniaxial compression of an assembly of lens-shaped particles under fixed confining pressure of 20kPa, showing rapid degradation of the plateau stress in the large strain regime.  For successive compression tests, all particles were re-used, but poured anew into the confining membrane so as to generate the same initial conditions for each test.  \textbf{(b)} Angle of internal friction versus compression test number for tetrahedra, spheres, and two different sets of lenses.}

\end{figure*}

While particle shape drives how contacting surfaces meet, the nature of the surfaces themselves tells the rest of the story.  
Many mechanical properties of granular materials have been found to depend strongly on the surface roughness of the particles \cite{SchroterFluidized2005,PohlmanTumblerRoughness2006,UtermannEtching2011,JanssenRoughness2011,TumblerPRE2016,GillemotSegregation2017}.
The process of plastic deformation erodes contacting surfaces and causes the properties of the granular material as a whole to evolve.
In geology, faults and the gouge between them scrape against one another, transforming the shape of the grains, their size dispersion, and the roughness of all surfaces over time \cite{FaultsSmooth2011,Renard2012RoughnessEvo}. 
In railroad engineering, wear and degradation undergone by the ballast beneath the rails results in an overall loss of shear strength, necessitating frequent maintenance and replacement \cite{RailroadDegradation2002}.  

Here we study the intertwined roles of particle shape and surface properties in a granular material by way of one particular shape's propensity to wear down under repeated shear.
We utilize 3D-printing to create thousands of identical particles and employ automated surface metrology to measure statistically representative samples of particle surfaces as an increasing number of compression tests are performed.
Comparing assemblies of particles, we find a striking change in behavior for convex lens-shaped particles and only marginal evolution for sphere and tetrahedron particles.  
From this we infer that the macro-scale behavior of the granular material can dramatically magnify changes that occur at the micro-scale, especially for particles whose dominant mechanism of relative motion under shear is sliding.
In particular, we show by combining x-ray tomography with high-resolution surface metrology that the degradation of strength of the lens particles is correlated with small changes in the particles’ surface roughness, while the overall particle shape and the distribution of contact locations on the particles’ surface remain essentially the same.

\section{Methods}
All particles were 3D-printed from UV-cured hard plastic (Young's modulus $E_{\text{mat}} \sim 1$GPa) using an Objet Connex 350 printer with a resolution of $\sim30\mu$m.  
Each particle's volume was 22.5mm$^3$. 
The lens particles used here were composed of the intersection of two spheres such that their surfaces meet at a latitude of 45 degrees.
The short and long axes of a lens particle were 2.1 and 5.1mm, respectively.

For each uniaxial compression test, approximately five thousand particles were poured randomly into a cylindrical, 0.6mm thick latex membrane, 5.0cm in diameter, to form a column with aspect ratio 2:1 (height to diameter). 
A vacuum pump evacuated the interior of the column, applying an isotropic confining pressure $\sigma_{\text{conf}} = 20$kPa. 
Uniaxial compression tests were performed on an Instron 5800 series materials tester under displace\-ment-controlled loading conditions. 
The tests were run with a strain rate of $\dot{\epsilon}=8\times10^{-4}$ s$^{-1}$ (0.05 min$^{-1}$), where the strain $\epsilon$ is the fractional axial displacement relative to the uncompressed column height. 
For successive compression tests, all particles were re-used, but poured anew into the confining membrane so as to generate the same initial conditions for each test.

To quantify the shear strength of the bulk granular material, we use the angle of internal friction $\psi$, calculated as in \cite{StressFlucsKM} from stress-strain data in the regime where steady-state plastic deformation occurs ($0.1\leq\epsilon<0.2$).  
We associate strains greater than 0.1 with a regime referred to in soil mechanics as the critical state \cite{SchofieldCriticalState1968}, where stress levels off and fluctuates around a mean value as the packing restructures via nonaffine, dissipative particle rearrangements
As the lens particles wear down, the assembly does not necessarily reach a constant plateau stress by the end of compression.  
Nevertheless, an average of the stress values in the regime $0.1\leq\epsilon<0.2$ gives a reproducible measure of the degradation of residual shear strength.  

In classic soil mechanics \cite{SchofieldCriticalState1968}, the angle of internal friction is calculated as
\begin{equation}
\text{sin}\psi = \frac{\sigma_3 - \sigma_1}{\sigma_3 + \sigma_1} = \frac{\bar{q}}{\bar{q}+2\sigma_{\text{conf}}}
\end{equation}
Here $\sigma_3$ and $\sigma_1$ are the largest and smallest principal stress components, equal to $\bar{q}+\sigma_{\text{conf}}$ and $\sigma_{\text{conf}}$ respectively, and $\bar{q}$ is the average deviatoric stress.
To correct for the effect of occasional large stress drops (see Fig. 1a) that bring the stress far below what is needed to cause plastic deformation, we calculate the plastic stress as the average of all of the stress values immediately preceding a stress drop \cite{StressFlucsKM}.

The rubber membrane also applies transverse confining stress due to its circumferential stretching as the packing expands outward during compression.  
This correction was calculated by Henkel and Gilbert \cite{HenkelMembrane1952} to be approximately linear in strain; for our membrane the effect is an additional 1.4kPa at $\epsilon=0.1$ and 2.8kPa at $\epsilon=0.2$, adding directly to $\sigma_1$.

\begin{figure}
  \includegraphics[width=\columnwidth]{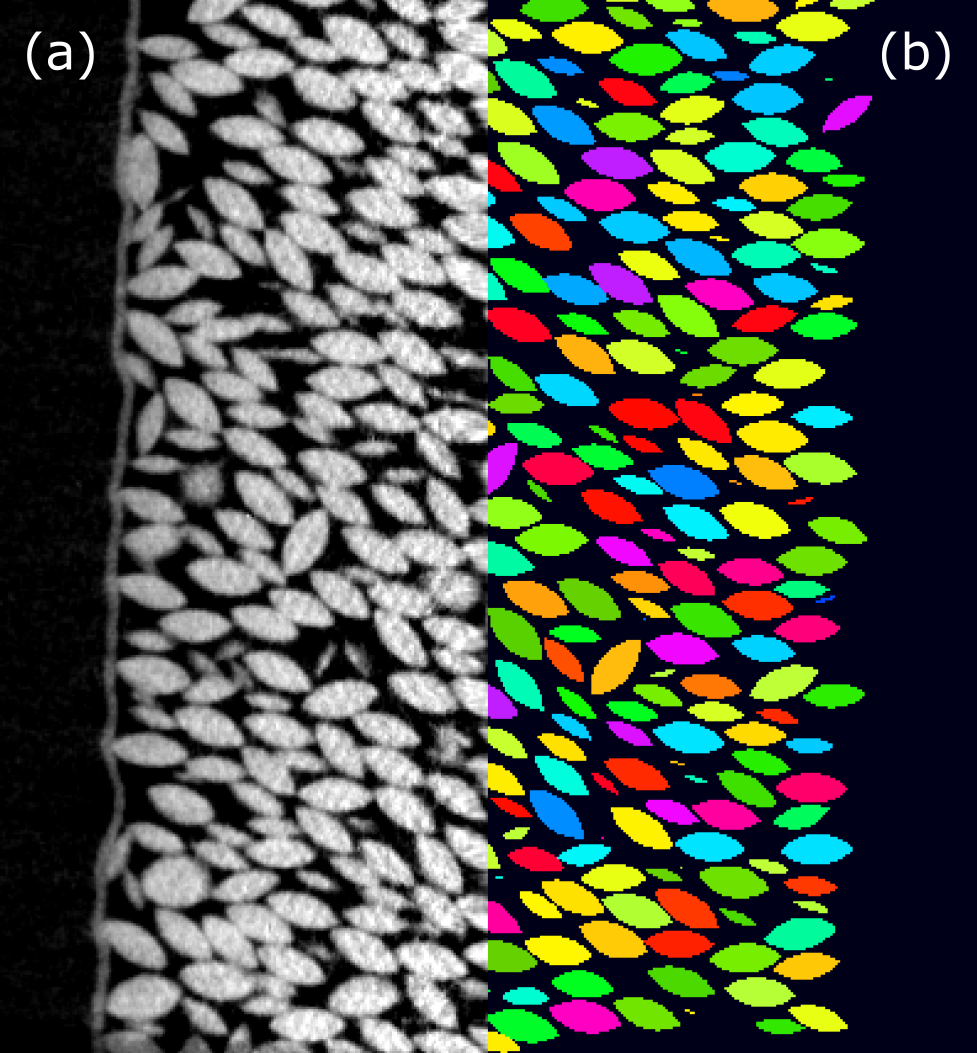}
\caption{Vertical slice through the center of an assembly of lens particles. \textbf{(a)} Raw tomography. \textbf{(b)}Reconstructed lenses. Panel \textbf{(a)}  shows the left half of the imaged slice, while \textbf{(b)} shows the right half of the same slice.}
\end{figure}

\subsection{Computed Tomography}

Computed tomography scans were taken with a C-arm x-ray scanner that we interfaced with the Instron materials tester, as described and characterized in \cite{ThanasiT}. 
The reconstruction of the total (6cm)$^3$ imaged volume was performed with voxels of 150$\mu$m side length;  as such each particle was represented by several hundred voxels.

Following techniques used to segment CT scans of granular material \cite{AsteSphereContacts,SchroterEllipsoidContacts,SchroterTets2013}, the raw tomography volume (Fig. 2a) was binarized and then segmented using a watershed algorithm.
This successfully segmented $\sim$85\% of the lens particles, which were then individually fit by numerical optimization that maximized overlap between the segmented volume and an ideal lens shape, as done with tetrahedra in Ref.\cite{SchroterTets2013}.
The fitted particles were then subtracted from the binarized tomography.
The remaining, initially unresolved portions of the tomography contained less distinct lens particles, which were iteratively fit and then subtracted individually.

To exclude boundary effects arising from the confining membrane, lenses within two particle widths were excluded from our analysis.
This made the region of interest a 3cm diameter cylindrical core in the center of the packing, inside which more than 99.5\% of the particles were successfully fit.  
The reconstructed lenses can be seen in Fig. 2b, which aligns seamlessly down the middle with the raw tomography.

Exact determination of contacts from a tomography of a granular packing is rendered impossible by limitations arising from a combination of factors that include particle imperfections, limited imaging precision together with uncertainties in reconstruction, and very generally the small length scales involved in defining an actual contact \cite{SchroterTets2013}.
The best we can do is to find physically justifiable criteria for identifying contacts which minimize the inclusion of `spurious' contacts and the exclusion of legitimate ones.
To this end we applied a procedure previously used with spheres \cite{AsteSphereContacts}, ellipsoids \cite{SchroterEllipsoidContacts,EllipsoidTomo2014}, and tetrahedra \cite{SchroterTets2013}.
To identify contacts in the packing, all particles were grown by uniform dilation, starting from the fitted centers and orientations, while counting the total number of contacts.
In the ideal case, i.e., in the absence of tomography noise, fitting errors, and particle imperfections, the average contact count across all particles in the assembly should jump discontinuously from zero to a nonzero value when the particles have been dilated to their actual size.
More realistically, the locations of the contacts can be assumed to be afflicted by noise from a normal distribution whose width folds in the various sources of error in the fitting process \cite{SchroterEllipsoidContacts}.
As a result, the measured contact count is expected to grow with the virtual particle length scale as a cumulative normal distribution with finite width $\sigma$.
Three full reconstructions were performed, both before and after load cycling and for different values of strain.
For each the value of $\sigma$ was less than 2\% of a particle diameter and approximately the same as in Ref. \cite{SchroterTets2013}, indicating a satisfactory fit of the tomography data.
From these reconstructions, spatial contact density distributions were obtained, as shown in Fig. 3.

\begin{figure}
  \includegraphics[width=\columnwidth]{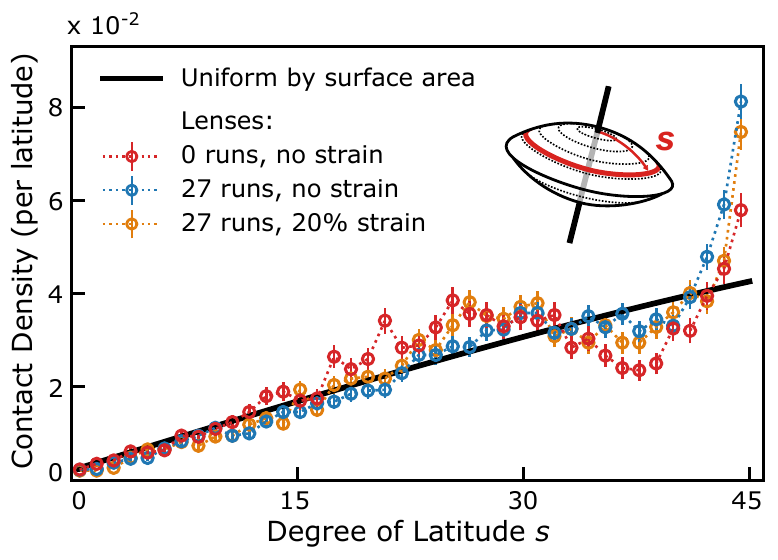}
\caption{Spatial contact density distributions for lens assemblies. Contact points on every particle are binned by degree of latitude, from 0 (the pole) to 45 (the edge). The distributions are normalized to integrate to unity, and the black line is the distribution that would result from contacts distributed uniformly across the surface area of a particle.  Data from three separate experiments are shown, with number of compressions and applied strain as indicated. }
\end{figure}

\subsection{Surface Metrology}
We used a LEXT OLS5000 laser confocal microscope to quantify changes in the micro-scale surface features of the particles. 
Immediately after 3D printing and then after every nine compression tests, 50 particles were removed from the granular material to be scanned from the top and 200 were removed to be scanned from the side.  
The scanned surface area constituted about 0.5\% of the total surface area of all particles in the packing, while the scanned edges were about 1\% of the total edge length, making the scans a small but statistically representative sample. 

Scans were taken with the 5$\times$ objective and 20$\mu$m vertical steps, yielding height map data with 2.5$\mu$m spacing between data points in the horizontal plane and a nominal resolution of 30nm in height.

Each data point of the height map was placed in its particle coordinate frame by numerically fitting an ideal lens shape to the entire scan, thus removing the overall particle curvature. 
Displacements normal to the ideal particle surface were calculated for each point, and the root mean square (rms) variation was used as a measure for roughness.

This spatial roughness calculation corresponds to the parameter $S_q$ in surface metrology \cite{blat}, calculated as 
\begin{equation}
    S_q = \sqrt{\langle(z-\bar{z})^2\rangle}.
\end{equation}
Here $\bar{z}$ is the average normal displacement and the brackets indicate an average over the surface.  We employ $S_q$ for both lens body and edge, where the edge roughness is extracted from a 50$\mu$m wide strip along the equator in each scan.

\section{Results \& Discussion}

In Fig. 1a the stress needed to plastically deform an assembly of lens particles has dropped by more than a factor of four after 27 compression cycles, and the angle of internal friction, shown in Fig. 1b, has decreased by almost 25 degrees, such that by 20 cycles the lenses support less shear stress than spheres.  
This evolution of the lens particles lies in stark contrast to the tetrahedra and spheres, for which after 19 and 30 compression tests, respectively, the angle of internal friction decreased only marginally from its starting value.  

Computed tomography scans of the lens particles provide hints at the origin of this rapid evolution with load cycling.  
Neighboring lens particles frequently contact each other in a brick-laying pattern, seen qualitatively in Fig. 2 and quantitatively in the excess contacts partway down the side of the particle body, shown in Fig. 3.
Because the smoothly sloping face of the lens lacks obstructing features, contacts midway down the body of the particle draw their resistance to shear predominantly from surface friction between particles.  
There is an additional abundance of contacts at the edges of particles, suggesting that the sharp equator of the lens shape also plays an outsized role in supporting shear stresses.
The lens shape's oblateness, which leads to the stacked particle ordering, causes a strong resistance to rotation about the particle's major axis, as this would require a high degree of local dilation.  
Because of this, the particles are more likely to slide relative to one another than to roll \cite{JensenDEMClusters,StannariusShape2013}.
This leads to increased grinding of the surfaces in contact and places more importance on the surface roughness when accounting for the assembly's overall resistance to shear \cite{AzemaPlaty2013}.
By contrast, tetrahedra and spheres would be more likely to roll in response to shear since their rotation dilates the surrounding volume less.
In turn, this would slow the pace of smoothing on the faces of the particles and lead to a diminished dependence of shear strength on surface roughness.

No significant changes in the contact distributions are found between particles before and after 27 compression runs, nor between the start and end of a compression test (Fig. 3).
That the contact distribution changes so little between as-poured fresh and worn particles suggests that the evolution of the strength of the lens assembly is not occurring at the mm-scale and is not a direct result of how the particles pack together to support shear stresses.  
Additionally, the lack of significant change in the contact distribution between as-poured assemblies and assemblies after 20\% applied strain indicates there is no disruptive restructuring of the contact network during compression, which could cause the drop in shear strength.

\begin{figure}
  \includegraphics[width=\columnwidth]{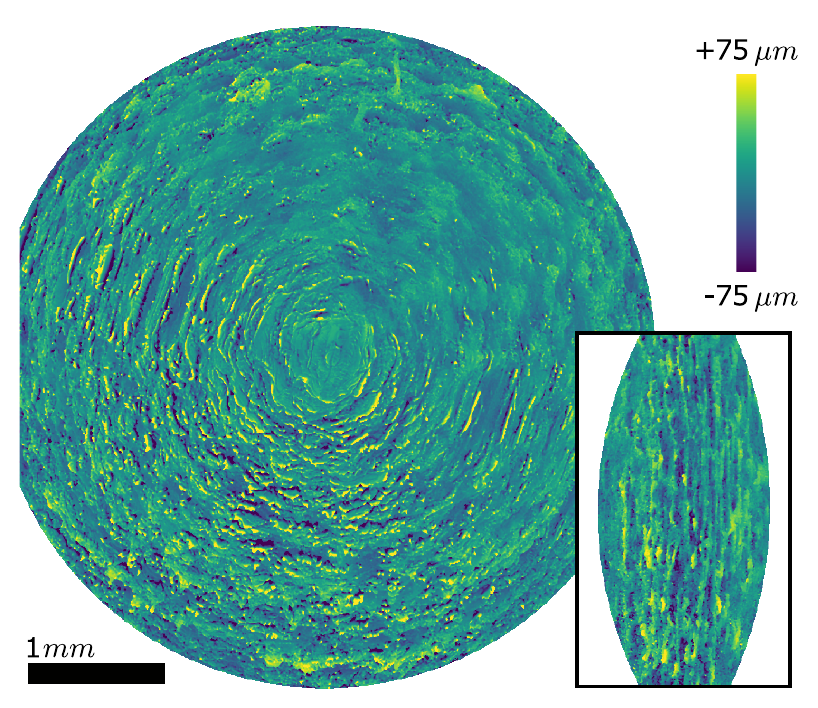}
\caption{Height deviations in the surface normal direction for a lens particle fresh after 3D printing viewed from the top (\textit{main}) and from the side (\textit{inset}). Both are displayed at the same scale}
\end{figure}

While the tomographic data indicate sliding, and thus particle surface friction, as the primary agent of shear strength for lens particles, the limited resolution of the reconstructed assemblies prohibits a direct analysis of the particle surface.
To probe these smaller length scales and directly quantify the degree of surface degradation, high-resolution optical surface scans were used.

These scans immediately show imperfections in the 3D printing process that engender a rugged particle surface (Fig. 4).  
Clearly apparent on the particle face at this magnification are grooves between print material layers in addition to accumulations of material deposited during the printing and curing process.  
We hypothesize that the jaggedness of these surface features is the main source of friction between the particles, and that any smoothing of these features would lead to lower friction and thus lower shear strength of the assembly as a whole.  
Further, we reason that the relative sliding between particles, driven by the plastic deformation of the lens assembly under compression, causes these features to grind and wear down gradually.

\begin{figure}
  \includegraphics[width=\columnwidth]{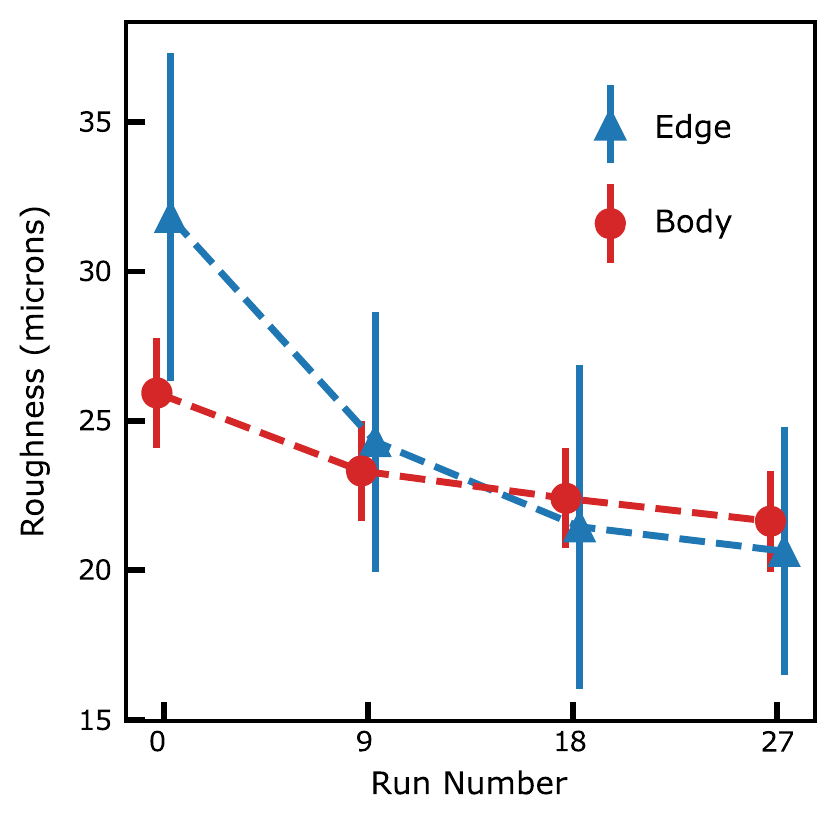}
\caption{Root mean square surface roughness $S_q$  for both lens edge and body as a function of number of compressions. Error bars show the standard deviation of the distribution of values in each case.  Data points are offset slightly in the horizontal direction for visual clarity.}
\end{figure}

As demonstrated in Fig. 5, the roughness coefficient $S_q$ for lens particles indeed changes systematically with successive compression cycles.
$S_q$ on the body of the particles decreases by about 15\% after 27 compressions.
Near the edges the particle surfaces start out rougher, but smooth out to a similar $S_q$ value, undergoing a drop by about 33\%. 
To fully characterize surfaces at contact and link this to a coefficient of friction in a quantitative manner, more comprehensive analysis is required \cite{KouraWear1981,GadelmawlaRuff2002,Menezes2008}.
Nevertheless, the decline in roughness $S_q$ is clearly measurable and correlates with the degradation of shear strength of the assembly.

That edges smooth at a faster rate than the rest of the particle surface is likely due to the preponderance of contacts near the edge (Fig. 3) as well as the fact that regions with higher curvature tend to erode faster than regions with lower curvature, a phenomenon also used to model pebble erosion \cite{DurianPebble2006,DomokosPebbles2014}.
On the freshly-printed lens particles, the curvature is $\sim(5000\mu \text{m})^{-1}$ and constant everywhere but the edge, where it becomes two orders of magnitude larger to reach $\sim(50\mu \text{m})^{-1}$.   
Note, however, that similarly sharp edges and corners also exist in the tetrahedra.
Any smoothing out of these high-curvature features in the particle geometry, therefore, is likely not the primary reason for the loss of shear strength seen in Fig. 1.
Instead, the ability of the lenses to change from high to low angle of internal friction appears to be driven by decreasing surface friction.
This, together with the oblate lens shape, favors sliding.
As Fig. 1b demonstrates, once the lens surfaces have become sufficiently smooth, such sliding can lead to an angle of internal friction or, equivalently, a shear strength of the granular assembly that is even lower than that for spheres.

\section{Conclusions}

The wear experienced by 3D printed particles under repeated compressive loading is found to drive systematic, controllable changes in their surface roughness. 
A most striking outcome is how the sensitivity to surface roughness depends on particle shape. 
The geometry of the convex lenses strongly imposes a degree of global ordering, shown in the tomography data (Fig. 2), whereby the particles tend to align their minor axes with gravity.  
This alignment leads to an overabundance of particle contacts midway down the body of the lens and at the edge. 
The robustness of this spatial contact distribution, combined with the oblique nature of the particle contacts themselves, primes the assembly to deform predominantly by sliding motion. 
Repeated sliding then smooths out the particle surfaces and enhances the ability to slide.
Finally, our finding that assemblies composed of more compact, less oblate particle shapes such as spheres and tetrahedra exhibit significantly lower sensitivity than lenses to repeated load cycling suggests new strategies by which to design granular materials that either experience wear in a predetermined way, or generate a desired global trend over repeated plastic deformation cycles. 

Our results highlight the complementary roles of particle shape and particle surface properties in determining the macro- and microscale behavior of granular materials.
Combining measurements of overall material properties such as the strength under shear with x-ray tomography and particle surface metrology opens the door to performing detailed granular forensics. 
By batch-processing large numbers of particles with laser-scanning confocal microscopy we show that the history of deformation as preserved in the patterns of wear on the particle surfaces can be extracted in a statistically meaningful manner.
Our results also demonstrate the importance of characterizing the surfaces of 3D printed granular materials, especially when designing overall properties by way of particle shape \cite{Miskin2013,Roth2016,HeinrichReview2015}.

\begin{acknowledgements}
This work is dedicated to the memory of Bob Behringer. We thank M. Lim and A. Kline for insightful discussions, and J. Chizewer and E. Ferguson for preliminary experiments. This work was supported by the National Science Foundation through Grant CBET-1605075. L.R. acknowledges support from the Center for Hierarchical Materials Design (CHiMaD) and from the Army Research Office under Grant Number W911NF-19-1-0245.	

\end{acknowledgements}

\section*{Compliance with ethical standards}
The authors declare that they have no conflict of interest.

\bibliographystyle{spphys}

\begin{thebibliography}{10}
	\providecommand{\url}[1]{{#1}}
	\providecommand{\urlprefix}{URL }
	\expandafter\ifx\csname urlstyle\endcsname\relax
	\providecommand{\doi}[1]{DOI \discretionary{}{}{}#1}\else
	\providecommand{\doi}{DOI \discretionary{}{}{}\begingroup
		\urlstyle{rm}\Url}\fi
	
	\bibitem{ThanasiShapes}
	A.G. Athanassiadis, M.Z. Miskin, P.~Kaplan, N.~Rodenberg, S.H. Lee, J.~Merritt,
	E.~Brown, J.~Amend, H.~Lipson, H.M. Jaeger, Soft Matter \textbf{10}(1), 48
	(2014)
	
	\bibitem{StannariusShape2013}
	T.~Börzsönyi, R.~Stannarius, Soft Matter \textbf{9}(31), 7401 (2013)
	
	\bibitem{GravishStaples2012}
	N.~Gravish, S.V. Franklin, D.L. Hu, D.I. Goldman, Physical Review Letters
	\textbf{108}(20), 208001 (2012)
	
	\bibitem{Zpaper}
	K.A. Murphy, N.~Reiser, D.~Choksy, C.E. Singer, H.M. Jaeger, Granular Matter
	\textbf{18}(2), 26 (2016)
	
	\bibitem{AnthonyParticleCharacteristics2005}
	J.L. Anthony, C.~Marone, Journal of Geophysical Research: Solid Earth
	\textbf{110}(B8) (2005)
	
	\bibitem{Santamarinasoilshape2004}
	J.~Santamarina, G.~Cho, in \emph{Advances in geotechnical engineering: The
		skempton conference}, vol.~1 (Thomas Telford, London, 2004), vol.~1, pp.
	604--617
	
	\bibitem{AzemaPentagons2007}
	E.~Az{\'e}ma, F.~Radja{\"i}, R.~Peyroux, G.~Saussine, Physical Review E
	\textbf{76}(1), 011301 (2007)
	
	\bibitem{StressFlucsKM}
	K.A. Murphy, K.A. Dahmen, H.M. Jaeger, Physical Review X \textbf{9}(1), 011014
	(2019)
	
	\bibitem{SchroterFluidized2005}
	M.~Schr\"oter, D.I. Goldman, H.L. Swinney, Phys. Rev. E \textbf{71}, 030301
	(2005)
	
	\bibitem{PohlmanTumblerRoughness2006}
	N.A. Pohlman, B.L. Severson, J.M. Ottino, R.M. Lueptow, Phys. Rev. E
	\textbf{73}, 031304 (2006)
	
	\bibitem{UtermannEtching2011}
	S.~Utermann, P.~Aurin, M.~Benderoth, C.~Fischer, M.~Schr\"oter, Phys. Rev. E
	\textbf{84}, 031306 (2011)
	
	\bibitem{JanssenRoughness2011}
	R.~Back, Granular Matter \textbf{13}(6), 723 (2011)
	
	\bibitem{TumblerPRE2016}
	L.T. Sheng, W.C. Chang, S.S. Hsiau, Phys. Rev. E \textbf{94}, 012903 (2016)
	
	\bibitem{GillemotSegregation2017}
	K.A. Gillemot, E.~Somfai, T.~Börzsönyi, Soft Matter \textbf{13}, 415 (2017)
	
	\bibitem{FaultsSmooth2011}
	E.E. Brodsky, J.J. Gilchrist, A.~Sagy, C.~Collettini, Earth and Planetary
	Science Letters \textbf{302}(1), 185 (2011)
	
	\bibitem{Renard2012RoughnessEvo}
	F.~Renard, K.~Mair, O.~Gundersen, Journal of Structural Geology \textbf{45},
	101 (2012)
	
	\bibitem{RailroadDegradation2002}
	P.~Tolppanen, O.~Stephansson, L.~Stenlid, Bulletin of Engineering Geology and
	the Environment \textbf{61}(1), 35 (2002)
	
	\bibitem{SchofieldCriticalState1968}
	A.~Schofield, P.~Wroth, \emph{Critical state soil mechanics}, vol. 310
	(McGraw-Hill London, 1968)
	
	\bibitem{HenkelMembrane1952}
	D.J. Henkel, G.D. Gilbert, G\'eotechnique \textbf{3}(1), 20 (1952)
	
	\bibitem{ThanasiT}
	A.G. Athanassiadis, P.J.L. Rivière, E.~Sidky, C.~Pelizzari, X.~Pan, H.M.
	Jaeger, Review of Scientific Instruments \textbf{85}(8), 083708 (2014)
	
	\bibitem{AsteSphereContacts}
	T.~Aste, M.~Saadatfar, T.J. Senden, Physical Review E \textbf{71}(6), 061302
	(2005)
	
	\bibitem{SchroterEllipsoidContacts}
	F.M. Schaller, M.~Neudecker, M.~Saadatfar, G.~Delaney, K.~Mecke, G.E.
	Schröder-Turk, M.~Schröter, AIP Conference Proceedings \textbf{1542}(1),
	377 (2013)
	
	\bibitem{SchroterTets2013}
	M.~Neudecker, S.~Ulrich, S.~Herminghaus, M.~Schröter, Physical Review Letters
	\textbf{111}(2), 028001 (2013)
	
	\bibitem{EllipsoidTomo2014}
	C.~Xia, K.~Zhu, Y.~Cao, H.~Sun, B.~Kou, Y.~Wang, Soft Matter \textbf{10}(7),
	990 (2014)
	
	\bibitem{blat}
	F.~Blateyron, in \emph{Characterisation of Areal Surface Texture}, ed. by
	R.~Leach (Springer Berlin Heidelberg, Berlin, Heidelberg, 2013), pp. 15--43
	
	\bibitem{JensenDEMClusters}
	R.P. Jensen, P.J. Bosscher, M.E. Plesha, T.B. Edil, International Journal for
	Numerical and Analytical Methods in Geomechanics \textbf{23}(6), 531 (1999)
	
	\bibitem{AzemaPlaty2013}
	M.~Boton, E.~Az{\'e}ma, N.~Estrada, F.~Radja{\"i}, A.~Lizcano, Physical Review
	E \textbf{87}(3), 032206 (2013)
	
	\bibitem{KouraWear1981}
	M.M. Koura, M.A. Omar, Wear \textbf{73}(2), 235 (1981)
	
	\bibitem{GadelmawlaRuff2002}
	E.S. Gadelmawla, M.M. Koura, T.M.A. Maksoud, I.M. Elewa, H.H. Soliman, Journal
	of Materials Processing Technology \textbf{123}(1), 133 (2002)
	
	\bibitem{Menezes2008}
	P.L. Menezes, Kishore, S.V. Kailas, Sadhana \textbf{33}(3), 181 (2008)
	
	\bibitem{DurianPebble2006}
	D.J. Durian, H.~Bideaud, P.~Duringer, A.~Schröder, F.~Thalmann, C.M. Marques,
	Physical Review Letters \textbf{97}(2), 028001 (2006)
	
	\bibitem{DomokosPebbles2014}
	G.~Domokos, D.J. Jerolmack, A.. Sipos, k.~Török, PLOS ONE \textbf{9}(2),
	e88657 (2014)
	
	\bibitem{Miskin2013}
	M.Z. Miskin, H.M. Jaeger, Nature Materials \textbf{12}, 326 (2013)
	
	\bibitem{Roth2016}
	L.K. Roth, H.M. Jaeger, Soft Matter \textbf{12}(4), 1107 (2016)
	
	\bibitem{HeinrichReview2015}
	H.M. Jaeger, Soft Matter \textbf{11}(1), 12 (2015)
	
\end{thebibliography}

\end{document}